\documentclass[review]{elsarticle}
\usepackage{amssymb}
\usepackage{amsfonts}
\usepackage{amsmath}
\usepackage{breqn}
\usepackage{qtree}
\usepackage{hyperref}
\usepackage[ruled,vlined]{algorithm2e}

\journal{arXiv}

\begin{document}

\begin{frontmatter}

\title{Privacy-preserving feature selection: \\A survey and proposing a new set of protocols}

\author[mymainaddress]{Javad Rahimipour Anaraki\corref{mycorrespondingauthor}}
\cortext[mycorrespondingauthor]{Corresponding author}
\ead{j.rahimipour@utoronto.ca}

\author[mysecondaryaddress]{Saeed Samet}
\ead{ssamet@uwindsor.ca}

\address[mymainaddress]{Institute of Biomedical Engineering, University of Toronto, Toronto, ON M5S 3G9 Canada}
\address[mysecondaryaddress]{School of Computer Science, University of Windsor, Windsor, ON, N9B 3P4, Canada}

\begin{abstract}
Feature selection is the process of sieving features, in which informative features are separated from the redundant and irrelevant ones. This process plays an important role in machine learning, data mining and bioinformatics. However, traditional feature selection methods are only capable of processing centralized datasets and are not able to satisfy today's distributed data processing needs. These needs require a new category of data processing algorithms called privacy-preserving feature selection, which protects users' data by not revealing any part of the data neither in the intermediate processing nor in the final results. This is vital for the datasets which contain individuals' data, such as medical datasets. Therefore, it is rational to either modify the existing algorithms or propose new ones to not only introduce the capability of being applied to distributed datasets, but also act responsibly in handling users' data by protecting their privacy. In this paper, we will review three privacy-preserving feature selection methods and provide suggestions to improve their performance when any gap is identified. We will also propose a privacy-preserving feature selection method based on the rough set feature selection. The proposed method is capable of processing both horizontally and vertically partitioned datasets in two- and multi-parties scenarios.
\end{abstract}


\end{frontmatter}

\section{Introduction}
\label{Introduction}
Collecting and accumulating data in a systematic way, such as in datasets and data tables, for further processing, is very important for any organization, department and, in a broader view, to any country. A dataset is composed of several data tables, in which each table contains several columns that correspond to variables (also called features or attributes) and rows which represent records (also called samples or objects). In machine learning, each dataset contains only one data table, and dataset and data table concepts are used interchangeably. As is shown in Table \ref{dataset}, for each row of a dataset, different variables are measured and inserted into the provided cells. Later, all the collected data are processed for a variety of purposes using different statistical and mathematical methods. \mbox{Table \ref{dataset}} shows a portion of Haberman's Survival dataset adopted from the UCI repository of machine learning \cite{Lichman}. In this dataset, each row represents a single patient, and columns show the age, year of operation and the number of auxiliary nodes of each patient. The last column presents a class label for each patient to show whether they have survived for five years or longer (represented by 1) or not (represented by 2). 

\begin{table}[!h]
\caption{Partial View of Haberman's Survival Dataset}
\centering
\begin{tabular}{cccc}
\hline
\bf{Age}& \bf{Year}& \bf{Auxiliary nodes} & \bf{Class} \\\hline
$\vdots$ & $\vdots$ & $\vdots$ & $\vdots$\\
42 & 61 & 4 & 1\\
42 & 62 & 20 & 1\\
42 & 65 & 0 & 1\\
42 & 63 & 1 & 1\\
43 & 58 & 52 & 2\\
43 & 59 & 2 & 2\\
43 & 64 & 0 & 2\\
43 & 64 & 0 & 2\\
43 & 63 & 14 & 1\\
43 & 64 & 2 & 1\\
$\vdots$ & $\vdots$ & $\vdots$ & $\vdots$ \\\hline
\end{tabular}
\label{dataset}
\end{table}

Since the collected data can be categorized either as sensitive (e.g., medical, financial, military) or non-sensitive (e.g., publicly available data, the UCI datasets) data, the algorithms applied should be selected accordingly, so that the data can be accessed only as is appropriate. With the dramatic increase in the amount of information generated annually, privacy challenges are becoming a serious issue for governments and health related organizations. Therefore, many countries are investing heavily in designing, implementing and applying privacy-preserving methods \cite{SRI}.

In US law, \emph{privacy} is the right ``to be let alone'' \cite{Cooley} and should be protected by taking proper actions \cite{Warren}. In computer science, \emph{privacy} of individuals deals with deciding how one's information will be used. For instance, someone's health information should be kept secure and be shared only with physicians who have been chosen by the patient. These concerns necessitate a category of data-mining methods called privacy-preserving data-mining. ``Privacy-preserving data-mining'' refers to knowledge extraction techniques specific to privacy criteria. The main goal of these processes is to introduce a trade-off between accuracy and the amount of information revealed publicly. Generally speaking, the amount of raw data produced is much greater than the information that needs to be extracted from them. Therefore, more efforts and time are needed to process, save and maintain those data for later processing (such as classification or clustering). Many problems in machine-learning, data-mining and pattern recognition involve big datasets. A high-dimensional dataset (e.g., DNA microarray data), in terms of number of features and samples, requires a huge effort to be processed. Therefore, feature selection (FS) methods are used to effectively reduce the size of datasets (in one direction) by selecting only the most relevant columns. These methods select the most informative features, which are highly correlated with the outcome and loosely dependent on other features, so as to minimize further processing. Since the size of datasets can also be reduced in terms of number of samples, sample selection (SS) methods have emerged to reduce the size of datasets by removing irrelevant samples. By employing FS and SS methods, dataset dimensions can be lowered and further processing can become more efficient. 

In this paper, existing feature selection algorithms which consider privacy concerns as well as the application in distributed datasets will be investigated. We will also perform a thorough comparison from different aspects, such as performance, applicability, security and privacy. The rest of this paper is organized as follows: Section 2 depicts background and Section 3 surveys related work. Section 4 discusses the proposed approaches and Sections 5 concludes the paper.

\section{Background}
Vast amounts of research have been conducted in different areas of data-mining and machine-learning to satisfy the need for protecting individuals' privacy \cite{Lindell,Agrawal,Samet}. Surprisingly however, feature selection methods have not kept up with the developing need for privacy and security. Feature selection is the process of purifying data by retaining the most informative features while omitting the others. The important role of feature selection methods in reducing model complexity for further processing is undeniable. Each dataset contains three types of features: informative, redundant and irrelevant. The most informative, non-redundant relevant features convey sufficient amounts of information for the outcome. Redundant features contain chunks of information that are indistinguishable from other similar features and can be removed. Features belonging to the last type are unnecessary (such as a feature with constant value for all examples) and can be eliminated due to not having any information for the classification outcome.

\subsection{Feature Selection}
Before looking at privacy-preserving aspects of data-mining, we will review some of the existing feature selection methods. Feature selection methods have been divided into two main groups: feature ranking and feature subset selection \cite{Hall}. The former is a set of methods that rank features based on some specific measure values and select the top $n$ number of features. The latter evaluates subsets of features and selects the one with the highest fitness value. Either of the aforementioned groups can be addressed using filter-based or wrapper-based approaches \cite{Kohavi}. In the filter-based approach, a merit evaluates the quality of every feature regardless of its impact on the outcome, while wrapper-based approaches measure the effectiveness of features based on the results of already chosen classifiers. Wrapper-based methods are highly computationally-intensive and powerful in predicting the outcome compared to filter-based methods, which are faster but potentially inaccurate.

\begin{figure}[!h]
	\centering
	\includegraphics[scale=.4]{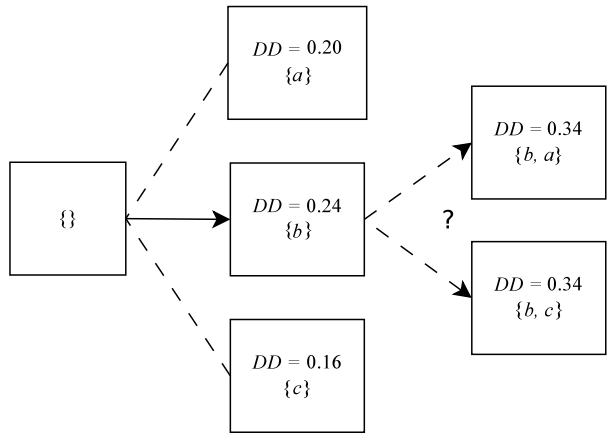}
	\caption{\emph{Equal} situation}
	\label{equal}
\end{figure}

One of the most well-known feature selection methods is Relief \cite{Kira}, which measures the relevancy of a feature compared to other features of the same and different classes by calculating their Euclidean distance. Hall \cite{Hall2} has proposed a merit based on the average intra-correlation of features and inter-correlation of features to the outcome. This merit selects features that are highly correlated to the outcome while lowly correlated to the other features. Jensen et al. \cite{Jensen} have introduced a novel feature selection method based on the lower approximation of a fuzzy-rough set, in which dependency of the features to the outcome is calculated using a merit called dependency degree (DD). Fuzzy-rough DD selects a new feature if it improves the discernibility power of the already selected features toward distinction of different classes of the outcome. Anaraki et al. \cite{Anaraki} have developed a simple control criterion for the conventional fuzzy-rough feature selection (FRFS) to direct the process of adding features to the reduct set by considering a lower bound for the distinguishability power of the feature being considered. Also, they have reviewed and surveyed different methods proposed in rough set feature selection (RSFS) in \cite{Anaraki2}. Anaraki et al. \cite{Anaraki3} have introduced the following two modifications of FRFS to improve the performance of the conventional method: guiding the selection process in \emph{equal} situations, where diverse subsets with only one different feature result in identical DD, and integrating the first improvement with the criterion that stops it \cite{Anaraki}. Figure \ref{equal} shows an \emph{equal} situation for subsets $\{b,a\}$ and $\{b,c\}$, in which the two sets differ by one member ($\{a\}$ and $\{c\}$) and for both $DD=0.34$ .

\subsection{Privacy-Preserving Data-Mining}

\begin{figure}[h!]
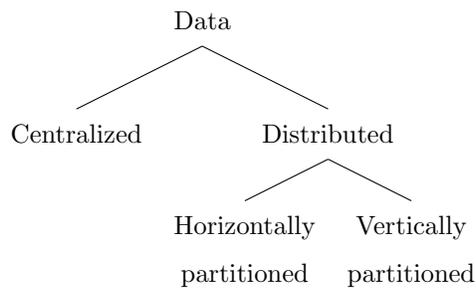

\Tree[.Data 
             Centralized
             [.Distributed 
                  [.{Horizontally\\partitioned} ]
                  [.{Vertically\\partitioned} ]
             ]
        ]
\caption{Data representation in privacy-preserving data-mining}
\label{datatree}
\end{figure}

There are two different approaches to privacy-preserving data-mining: methods to perturb data before publishing which are called randomization, and methods to perform mathematical operations securely which are called secure multi-party computation (SMC). Figure \ref{datatree} shows how data are represented to privacy-preserving data-mining methods.

\sloppy
In 2000, interestingly, two papers with the identical title of \emph{Privacy-Preserving Data-Mining} were published \cite{Agrawal,Lindell}. Agrawal and Srikant \cite{Agrawal} proposed a secure decision tree classifier, which can be applied to the perturbed and randomized data by reconstructing distribution using a Bayesian procedure. In the other paper, Lindell and Pinkas \cite{Lindell} proposed a secure protocol for the ID3 classifier on two-party horizontally partitioned data. The core of their method is a secure version of $x ln x$ in which $x$ is a two-party distributed data. Since the publication of these two seminal works, many protocols and methods have been proposed using both approaches for various data-mining, machine-learning and statistical analysis methods and algorithms.

\subsubsection{Centralized}
\paragraph{Randomization}
In this approach, data is centralized and the data owner wishes to publish their data for mining purposes. To do so, the data should be perturbed using randomization techniques before being transmitted, and are then reconstructed at the destination. The main challenge in randomization is the trade-off between privacy and accuracy.

Features in privacy-preserving data-mining are divided into three main categories: explicit identifiers (EI), quasi identifiers (QI) and sensitive identifiers (SI). Explicit identifiers are those features of a dataset which promptly reveal individuals' identification, such as name and medical care plan (MCP) number. Theses features should be removed to protect an individual's privacy. Quasi identifiers are those features which could be combined with publicly available data such as Netflix movies ranking to identify individuals. In 2006, Netflix released information on 100 million ratings to a competition called Netflix Prize to challenge researchers in order to find the best algorithm for predicting user ratings \cite{Bennett}. However, a few months later Netflix ratings were linked to the internet movie database (IMDB) ratings and individuals were identified \cite{Narayanan}. Sensitive identifiers refer to that information which is private to some individuals, such as disease information in medical datasets and should be also removed from the dataset~\cite{Aggarwal2015}.

In the case of having sensitive attributes, three methods have been proposed to protect individuals' privacy as follows:

\begin{enumerate}
\item {$k$-anonymity}: If each record in a dataset is indistinguishable from $(k-1)$ other records (see Table \ref{3-anonymized} adopted from \cite{Aggarwal2015})
\item {$l$-diversity}: If an equivalent class of a dataset has $l$ diverse values for the sensitive attribute
\item {$t$-closeness}: If the distance of the distribution of a sensitive attribute value in an equivalent class to the distribution of the same attribute is less than $t$
\end{enumerate}

\begin{table}[!h]
\caption{An example of 3-anonymized dataset}
\centering
\begin{tabular}{cccc}
\hline
\bf{Row Index} & \bf{Age} & \bf{ZIP Code}& \bf{Disease}\\\hline
$1$ & [20, 30] & Northeastern US & HIV\\
$2$ & [30, 40] & Western US & Hepatitis C\\
$3$ & [20, 30] & Northeastern US & HIV\\
$4$ & [30, 40] & Western US & Hepatitis C\\
$5$ & [30, 40] & Western US & Diabetes\\
$6$ & [20, 30] & Northeastern US & HIV\\\hline
\end{tabular}
\label{3-anonymized}
\end{table}

\subsubsection{Distributed}
\paragraph{Secure Multi-Party Computation}
In SMC, \emph{secure} mathematical and statistical computations are applied to different portions of data in the possession of different parties. This approach has the same results as non-secure algorithms; however, the main challenge in secure methods is the trade-off between security and efficiency.
In an $n$-party environment, datasets are divided into $n$ chunks and all parties demanding of running a specific mining algorithm (e.g., classification) or statistical analysis (e.g. correlation coefficient) on all $n$ chunks as a single dataset without revealing any private information to the others. 

\begin{figure}[h!]
	\centering
	\includegraphics[scale=.6]{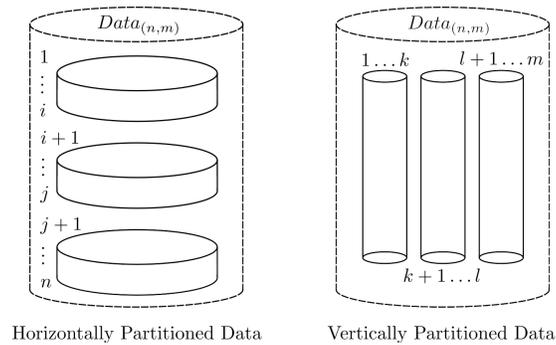}
	\caption{Horizontally and vertically partitioned data}
	\label{SMC}
\end{figure}

Data in SMC can be partitioned either \emph{vertically} or \emph{horizontally} (see Figure \ref{SMC}) and depending on how the data are partitioned, ``partition-specific'' methods need to be applied. It is worth mentioning that data partitioning here is different from database partitioning in the distributed database management system \cite{Beach}, in which the main goal is to improve performance. In \emph{vertical} partitioning, each party (e.g., different departments of a store) might posses a subset of features (e.g., purchased items from a specific department) while accommodating all samples (e.g., customers). In \emph{horizontal} partitioning, each party might posses a subset of samples while accommodating all features. For example, Hospital A in Newfoundland and Labrador, Hospital B in Ontario and Hospital C in British Columbia have a Haberman's Survival dataset (as shown in Table \ref{dataset}) of people in their provinces. So, they all share the same structure and features for their datasets, but different records. 

\section{Related Work}
In this section we will discuss three privacy-preserving feature selection methods which have been introduced in the last seven years. Each method is presented in detail and in case of identifying any gap, some comments are provided to improve the performance of the proposed methods.

\begin{figure}[h!]
	\centering
	\includegraphics[scale=.59]{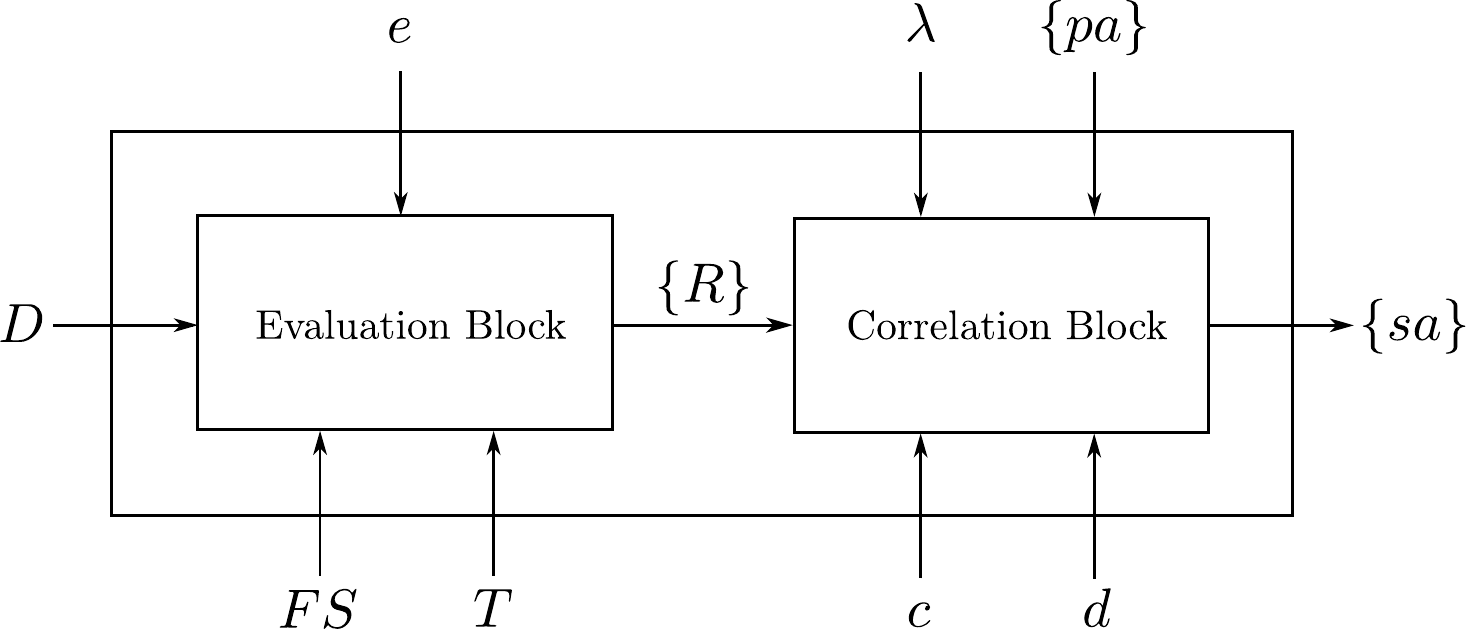}
	\caption{Privacy-aware filter-based feature selection}
	\label{JaferFig}
\end{figure}

Jafer et al. \cite{Jafer} have proposed a privacy-aware filter-based feature selection that probes the inter-correlation of features to remove quasi-identifier (QI) features. In their paper, the authors introduce a system which contains two separate blocks: one for evaluating features, and the other one for controlling the privacy aspects of feature selection. In the former, features are ranked based on InfoGain \cite{Hall3} and Relief criteria \cite{Kira}. In the latter, the list is traversed from bottom to top and correlation of QI features and non-QI features is calculated. By referring to the controlling values of the Correlation Block, features are selected or discarded. We have adopted Figure \ref{JaferFig} from \cite{Jafer} to illustrate the proposed system.

The Evaluation Block accepts dataset $D$, ranker threshold $T$, evaluation measure $e$, such as information gain and chi-square, and the type of feature selection $FS$, such as feature subset as inputs and $\{R\}$ is the ranked list features in the intermediate output. Later, the Correlation Block takes the ranked list $\{R\}$, the privacy-breaching attributes $\{pa\}$, the correlation measure $c$, such as symmetric uncertainty, the discretization factor $d$ for converting continuous values of the features to discrete ones as the requirement of correlation measure, and the balancing parameter between privacy and accuracy $\lambda$, in order to produce the set of selected attributes $\{sa\}$ as the final output. The balancing parameter $\lambda$ varies from zero to one, in which moving from zero to one increases the number of the selected QI features. 

After evaluating features using $e$, features are sorted in descending order list. Then, the correlation of the QI features in the list is calculated from bottom to up against other features using $c$, and those QI features which have a correlation greater than $\lambda$ will be removed. This process repeats until all QI features are investigated. This method guarantees to preserve privacy and select the most important features; however, there are four concerns about the method as follows:

\begin{enumerate}
\item Inter-correlation of the features and the class have not been investigated
\item In case of having two perfectly correlated QI and non-QI features, only the QI feature is removed
\item The case where more than one perfectly correlated non-QI feature to a QI feature exists has not been discussed
\item Security and complexity analysis of the proposed method are missing
\end{enumerate}

Banerjee and Chakravarty \cite{Banerjee} have developed a distributed privacy-preserving method based on the virtual dimensionality reduction method in image processing \cite{Chang} to select features. This method takes advantage of correlation and covariance eigenvalues to perform feature selection on both horizontally and vertically partitioned data. It starts with calculating the correlation and the convenience matrices of a dataset, and continues with computing the eigenvalues of both of the matrices. Then, for each feature, the corresponding correlation-eigenvalue and covariance-eigenvalue is subtracted. If the resulting value is greater than a user-specified threshold $\delta$ then the feature is kept in the reduct subset, otherwise, it will be discarded. This process continues until all features of the dataset are examined.

For both horizontally and vertically partitioned data, the correlation and the covariance are calculated securely over all parties; however, the eigenvalue decomposition is done locally to reduce communication costs. In both scenarios, the threshold $\delta$, number of rows of dataset $D$ as $N(D)$, summation of feature $j$ values as $FS_j(D)$, standard deviation of each feature $\sigma(j)$, and sum of product of values of features $i$ and $j$ as $SS_{ij}(D)$ should be calculated. Finally, the covariance and the correlation between the two features $i$ and $j$ are computed as shown in Equation \ref{cov} and \ref{corr}, respectively.

\begin{equation} 
COV(i,j) = \frac{SS_{ij}(D)}{N(D)} - \frac{FS_i(D) \times FS_j(D)}{N(D)^2},
\label{cov}
\end{equation}

\begin{equation} 
CORR(i,j) = \frac{COV(i,j)}{\sigma(i) \times \sigma(j)}.
\label{corr}
\end{equation}

For horizontally partitioned data, each party $p$ calculates $N^p(D)$, $FS^p_j(D)$, $\sigma(j)^p$ and $SS^p_{ij}(D)$, and then they apply secure sum protocol to calculate the aggregation results. Finally, each party performs feature selection based on the resulting eigenvalues of the calculated covariance and the correlation. With $n$ number of records, $m$ features and $p$ parties, the communication cost would be $O(mp)$, since the only secure operation used in the horizontally partitioned data is secure sum.

For vertically partitioned data, each party can calculate $N(D)$, $FS_j(D)$, $\sigma(j)$ locally; however, the calculation of $SS^p_{ij}(D)$ depends on whether both attributes of $i$ and $j$ are in the same partition or not. If so, the calculation is straight forward. Otherwise, both parties should use secure dot product to calculate $SS_{ij}(D)$. The communication cost of the vertically partitioned data is $O(m^2np)$, which is mainly imposed by the secure dot product operation.

Das \cite{Das} et al. have introduced three asynchronous feature selection methods based on the misclassification gain, Gini index and entropy measures for binary-class datasets with categorical features in horizontally partitioned fashion. The main requirement for the proposed methods is a P2P network with a structured ring-based topology. The distributed setup of each measure (i.e. misclassification gain, Gini index and entropy) to evaluate every feature $A_i$ are shown in Equations \ref{misclassification}, \ref{Gini} and \ref{entropy}, respectively. 

\begin{equation}
\sum_{a=0}^{m_i-1} \left | \sum_{l=1}^{d} \left\{ x_{i,a0}^{(l)} - x_{i,a1}^{(l)}  \right\} \right |,
\label{misclassification}
\end{equation}
where each feature $A_i$ can take a value from $\{0, \ldots, m_i -1 \}$, $d$ is the number of peers, and $x_{i,a0}^{(l)}$ and $x_{i,a1}^{(l)}$ are the number of examples with the value of $A_i = a$ and class value of $0$ and $1$, respectively.

\begin{equation}
\sum_{a=0}^{m_i-1} \left\{ \frac{ \left (  \sum_{l=1}^{d} x_{i,a0}^{(l)} \right )^2 +  \left (  \sum_{l=1}^{d} x_{i,a1}^{(l)} \right )^2 } { \sum_{l=1}^{d} x_{i,a}^{(l)}}   \right\},
\label{Gini}
\end{equation}
where $x_{i,a}^{(l)}$ is the number of examples with $A_i = a$.

\begin{equation}
\sum_{a=0}^{m_i-1} \left\{ \left (  \sum_{l=1}^{d} x_{i,a0}^{(l)} \right ) \log \left (\frac{\sum_{l=1}^{d} x_{i,a0}^{(l)} }{ \sum_{l=1}^{d} x_{i,a}^{(l)} } \right ) + \\  \left ( \sum_{l=1}^{d} x_{i,a1}^{(l)}  \right ) \log \left ( \frac{\sum_{l=1}^{d} x_{i,a1}^{(l)} }{\sum_{l=1}^{d} x_{i,a}^{(l)}}  \right ) \right\}.
\label{entropy}
\end{equation}

For computing misclassification gain, Gini index, and entropy across all peers, each peer $P_i$ estimates Equation \ref{misclassification}, \ref{Gini}, and \ref{entropy} for feature $A_i$ when it takes value $a$, respectively. This process starts from an initiator and continues by each peer with adding their value to the received data. When the initiator receives the data, it calculates the average using the asymmetric network topology version (see Equation \ref{avg}) of the method proposed by Scherber and Papadopoulos \cite{Scherber}.
\begin{equation}
z^{(t)}_i = \{1-2 \rho |\Gamma_{i,1}| - \rho(n^*_i - |\Gamma_{i,1}|)\} z_i^{(t-1)} +\\ 2 \rho \sum_{q \in \Gamma_{i,1}}z_q^{(t-1)} + \rho \sum_{q = 1}^{n_i^* - |\Gamma_{i,1}|} z_q^{(t-1)},
\label{avg}
\end{equation}
where $z^{(t)}_i$ is an estimate of average at the time $t$ by $i$th peer, $\rho$ is the rate of convergence, $|\Gamma_{i,1}|$ is the size of set of neighbours of peer $i$ in hop-distance one, and $n_i^*$ is the size of the ring formed by peer $i$.

To establish a trade-off between privacy and the cost of computations, the authors have introduced an objective function for each peer $i$ as follows:
\begin{equation*}
f^{\text{obj}}_i = w_{ti} \times \text{threat} - w_{ci} \times \text{cost},
\end{equation*}
where $w_{ti}$ and $w_{ci}$ are the weights for \emph{thread} and \emph{cost}, \emph{threat} is a measure which represents the risk that each peer might take by participating in the current computation, and \emph{cost} includes both computation and communication costs. The time complexity for the proposed methods based on three measures is $O(max(n^*_i, n^*_j))$, where $n^*_i$ is the optimal value for peer $P_i$ and $n^*_j$ is the value for the neighbour $P_j$ in the same ring. 

In this section, we have discussed three privacy-preserving feature selection methods, each trying to address feature selection issues in distributed datasets through decentralized computation.

\section{Discussion and Contribution}
All the discussed methods have provided a variety of secure protocols for different data configurations (i.e. horizontally and vertically partitioned data) to preserve individuals' privacy. The proposed methods have been backed up with security, computational and communication complexity analyses. However, there are obvious limitations for the mentioned privacy-preserving feature selection methods from privacy-preserving aspects. On one side, they can be applied to either horizontally or vertically partitioned data which limits users by imposing specific data configuration. On the other side, they are not suited for both two- and multi-party data configurations. 
To address the mentioned deficiencies we are providing four privacy-preserving versions of rough set feature selection \cite{Komorowski} for both horizontally and vertically partitioned datasets as follows:

\begin{itemize}
\item Two-party with horizontally partitioned data (2P-HP)
\item Multi-party with horizontally partitioned data (MP-HP)
\item Two-party with vertically partitioned data (2P-VP)
\item Multi-party with vertically partitioned data (MP-VP)
\end{itemize}

\section{Proposed method}
Rough set theory was proposed by Pawlak as a tool for dealing with uncertainty \cite{Pawlak}. Data in rough set theory are organized in a decision table. Table \ref{dt} shows a decision table adopted from \cite{Komorowski}. Class attribute is called decision attribute and the rest are condition attributes. In Table \ref{dt}, $\{Walk\}$ is a decision attribute and $\{Age, LEMS\}$ are condition attributes.

\begin{table}[!h]
\caption{An example of decision table}
\centering
\begin{tabular}{cccc}
\hline
\bf{Object} & \bf{Age} & \bf{LEMS} & \bf{Walk}\\\hline
$x_1$ & 16-30 & 50 & Yes\\
$x_2$ & 16-30 & 0 & No\\
$x_3$ & 31-45 & 1-25 & No\\
$x_4$ & 31-45 & 1-25 & Yes\\
$x_5$ & 46-60 & 26-49 & No\\
$x_6$ & 16-30 & 26-49 & Yes\\
$x_7$ & 46-60 & 26-49 & No\\\hline
\end{tabular}
\label{dt}
\end{table}

Let $\mathbb{U} = \{x_1, x_2,\ldots, x_7\}$ be the universe of discourse and let $R$ be the equivalence relation on $\mathbb{U}$, approximation space is shown by $(\mathbb{U},R)$. Set of all attributes are shown by $A = \{AGE, LEMS, WALK\}$, set of all conditional attributes by $C = \{Age, LEMS\}$ and set of decision attribute(s) or class attribute(s) by $D = \{WALK\}$. Let $X$ be a subset of $\mathbb{U}$ and $P$ to be a subset of $A$, approximating this subset using rough set theory is done by means of upper and lower approximations. Upper approximation of $X$ with regards to ($\overline{P}X$) contains objects which are possibly classified in $X$ regarding the attributes in $P$. Objects in lower approximation ($\underline{P}X$) are the ones which are surely classified in $X$ regarding the attributes in $P$. Boundary region of $X$ can be determined by subtracting upper approximation from lower approximation and where it is a non-empty set, $X$ is called a rough set otherwise it is a crisp set. Rough set is shown by an ordered pair ($\overline{P}X$, $\underline{P}X$). Different regions are defined using this pair as below:

\begin{equation}\label{eq:pos}
POS_{P}(Q) = \bigcup_{X \in \mathbb{U}/Q} \underline{P}X
\end{equation}

\begin{equation}\label{eq:neg}
NEG_{P}(Q) = \mathbb{U} - \bigcup_{X \in \mathbb{U}/Q} \overline{P}X
\end{equation}

\begin{equation}\label{eq:bnd}
BND_{P}(Q) = \bigcup_{X \in \mathbb{U}/Q} \overline{P}X - \bigcup_{X \in \mathbb{U}/Q} \underline{P}X
\end{equation}

Positive region of partition $\mathbb{U}/Q$ is a set of all objects which can be uniquely classified into blocks of partition by means of $P$. Negative region is a set of objects which cannot be classified to the partition $\mathbb{U}/Q$\cite{Komorowski}.

Finding dependency between attributes is one of the most important areas in data analysis. Let $P$ and $Q$ be subsets of $A$, dependency of $Q$ on $P$ are denoted by $P\Rightarrow_k Q$ and $k = \gamma_{p}(Q)$, in which $\gamma$ is dependency degree \cite{Komorowski}. If $k = 1$, $Q$ depends totally on $P$ and if $k < 1$, $Q$ depends partially on $P$.

Value of $k$ is a measure of dependency between features. In feature selection, those features which are loosely dependent on each other and highly correlated to decision feature are desired. If $Q$ totally depends on $P$, it means that the partition which is made by $P$ is finer than $Q$. Calculating dependency is shown in Equation \ref{eq:dd}.

\begin{equation}\label{eq:dd}
k = \gamma_{P}(Q) = \frac{|POS_{P}(Q)|} {|\mathbb{U}|}
\end{equation}

The notation $|.|$ is used for cardinality. Positive region of the partition $\mathbb{U}/Q$ with respect to $P$ which is denoted by $\gamma$, is the set of all elements which can be classified to partition $\mathbb{U}/Q$ using $P$ \cite{Komorowski}. Reduct is a subset of features which has the same dependency degree as employing all features for classification. Features which belong to the reduct set are information-rich and the others are irrelevant and redundant.

The QuickReduct algorithm which is given in \cite{Jensen} and depicted in Algorithm \ref{QR}, calculates a reduct without finding all the subsets. It starts from an empty set and each time selects a feature which causes greatest increase in dependency degree. The algorithm stops when adding more features does not increase the dependency degree. It does not guarantee to find minimal reduct as long as it employs greedy forward search algorithm, which is vulnerable to local optimum. 

\begin{algorithm}[!]
 \caption{QuickReduct algorithm\label{QR}}
$C$, the set of all conditional attributes\\
$D$, the set of decision attributes\\
$R \leftarrow\{\}; \gamma_{best}=0;\gamma_{prev}=0$\\
\bf do\\
\ \ $T \leftarrow R$\\
\ \ $\gamma_{prev}\leftarrow \gamma_{best}$\\
\ \ \bf foreach $x \in (C-R)$\\
\ \ \ \ \bf if ${\gamma_{R\cup \{x\}}(D)}>{\gamma_{T}(D)}$\\
\ \ \ \ \ \ $T\leftarrow R \cup \{x\}$\\
\ \ \ \ \ \ $\gamma_{best} \leftarrow \gamma_{T}(D)$\\
\ \ \ \ \ \ $R \leftarrow T$\\
\bf until $ \gamma_{best}=\gamma_{prev}$\\
\bf return $R$\\
\end{algorithm}

The QuickReduct algorithm has been applied to the example dataset in Table \ref{dt}. The algorithm starts by calculating dependency of the outcome $\{WALK\}$ to each conditional features $\{Age, LEMS\}$ as shown in Equation \ref{QR1}.

\begin{align}\label{QR1}
\nonumber
\gamma_{\{Age\}}(Walk) &= \frac{|POS_{\{Age\}}(Walk)|} {|\{x_1, x_2, x_3, x_4, x_5, x_6, x_7\}|} \\ \nonumber
&= \frac{|\bigcup_{X \in \{x_1, x_2, x_3, x_4, x_5, x_6, x_7\}/Walk} \underline{Age}X|} {|\{x_1, x_2, x_3, x_4, x_5, x_6, x_7\}|}\\ 
&= \frac{|{\{x_5, x_7\}}|} {|\{x_1, x_2, x_3, x_4, x_5, x_6, x_7\}|} = \frac {2}{7}
\end{align}
\begin{align*}
 \nonumber
\gamma_{\{LEMS\}}(Walk) &= \frac{|POS_{\{LEMS\}}(Walk)|} {|\{x_1, x_2, x_3, x_4, x_5, x_6, x_7\}|} \\ \nonumber
&= \frac{|\bigcup_{X \in \{x_1, x_2, x_3, x_4, x_5, x_6, x_7\}/Walk} \underline{LEMS}X|} {|\{x_1, x_2, x_3, x_4, x_5, x_6, x_7\}|}\\ \nonumber
&= \frac{|{\{x_1, x_2\}}|} {|\{x_1, x_2, x_3, x_4, x_5, x_6, x_7\}|} = \frac {2}{7} \nonumber
\end{align*}

Since, the dependency degree of $\{LEMS\}$ is equal to $\{Age\}$, either of them can be selected and added to the reduct set $R$. This process continues by selecting $\{LEMS\}$ and adding $\{Age\}$ to the reduct set; the dependency degree of the set is calculated as shown in Equation \ref{QR2}.

\begin{align}\label{QR2}
\nonumber
\gamma_{\{Age, LEMS\}}(Walk) &= \frac{|POS_{\{Age, LEMS\}}(Walk)|} {|\{x_1, x_2, x_3, x_4, x_5, x_6, x_7\}|} \\ 
&= \frac{|\bigcup_{X \in \mathbb{U}/Walk} \underline{Age, LEMS}X|} {|\{x_1, x_2, x_3, x_4, x_5, x_6, x_7\}|}\\ \nonumber
&= \frac{|{\{x_1, x_2, x_5, x_6, x_7\}}|} {|\{x_1, x_2, x_3, x_4, x_5, x_6, x_7\}|} = \frac {5}{7}\\ \nonumber
\end{align}

As the resulting dependency of having both features in the reduct set is greater than dependency degree of $R = \{LEMS\}$; therefore, the final result of QuickReduct algorithm is $R = \{Age, LEMS\}$.

\subsection{Two parties with horizontally partitioned data (2P-HP)}
As an illustrative example, Table \ref{dt} has been partitioned horizontally into two datasets $\mathbb{D}^H_1$ and $\mathbb{D}^H_2$ in possession of two parties $\mathbb{P}_1$ and $\mathbb{P}_2$, as shown in Tables \ref{P1} and \ref{P2}, respectively.

\begin{table}[!h]
\caption{The first partition of horizontally partitioned $\mathbb{D}^H_1$}
\centering
\begin{tabular}{cccc}
\hline
\bf{Object} & \bf{Age} & \bf{LEMS} & \bf{Walk}\\\hline
$x_1$ & 16-30 & 50 & Yes\\
$x_2$ & 16-30 & 0 & No\\
$x_3$ & 31-45 & 1-25 & No\\
$x_4$ & 31-45 & 1-25 & Yes\\\hline
\end{tabular}
\label{P1}
\end{table}

\begin{table}[!h]
\caption{The second partition of data $\mathbb{D}^H_2$}
\centering
\begin{tabular}{cccc}
\hline
\bf{Object} & \bf{Age} & \bf{LEMS} & \bf{Walk}\\\hline
$x_5$ & 46-60 & 26-49 & No\\
$x_6$ & 16-30 & 26-49 & Yes\\
$x_7$ & 46-60 & 26-49 & No\\\hline
\end{tabular}
\label{P2}
\end{table}

In order to uncover the required secure mathematical equations of 2P-HP for calculating dependency degree of each partitioned data $\gamma_{P}(Q)_{\mathbb{D}^H_1}$, the results of applying QuickReduct algorithm on each partition based on conditional feature $\{Age\}$, is calculated in Equation \ref{2P-HR-QR1}.

\begin{align}\label{2P-HR-QR1}
\nonumber
\gamma_{\{Age\}}(Walk)_{\mathbb{D}^H_1} &= \frac{|POS_{\{Age\}}(Walk)_{\mathbb{D}^H_1}|} {|\{x_1, x_2, x_3, x_4\}|} \\ \nonumber
&= \frac{|\bigcup_{X \in \{x_1, x_2, x_3, x_4\}/Walk} \underline{Age}X|} {|\{x_1, x_2, x_3, x_4\}|}\\ \nonumber
&= \frac{|{\{\}}|} {|\{x_1, x_2, x_3, x_4\}|} = \frac {0}{4}\\ 
\\ \nonumber
\gamma_{\{Age\}}(Walk)_{\mathbb{D}^H_2} &= \frac{|POS_{\{Age\}}(Walk)_{\mathbb{D}^H_2}|} {|\{x_5, x_6, x_7\}|} \\ \nonumber
&= \frac{|\bigcup_{X \in \{x_5, x_6, x_7\}/Walk} \underline{Age}X|} {|\{x_5, x_6, x_7\}|}\\ \nonumber
&= \frac{|{\{x_5, x_6, x_7\}}|} {|\{x_5, x_6, x_7\}|} = \frac {3}{3}\\ \nonumber
\end{align}

By referring to the resulting dependency degree of conditional feature $\{Age\}$ for partition ${\mathbb{D}^H_1}$ and ${\mathbb{D}^H_2}$; it can be understood that the overall dependency degree of conditional feature $\{Age\}$ cannot be calculated by simply adding corresponding numerator and denominator of $\gamma_{\{Age\}}(Walk)_{\mathbb{D}^H_1}$ to $\gamma_{\{Age\}}(Walk)_{\mathbb{D}^H_2}$ as shown in Equation \ref{2P-HR-Res}.

\begin{align}\label{2P-HR-Res}
\nonumber
\gamma_{\{Age\}}(Walk) &= \gamma_{\{Age\}}(Walk)_{\mathbb{D}^H_1} + \gamma_{\{Age\}}(Walk)_{\mathbb{D}^H_2} \\
&= \frac{|POS_{\{Age\}}(Walk)_{\mathbb{D}^H_1}|} {|\{x_1, x_2, x_3, x_4\}| + |\{x_5, x_6, x_7\}|} \\ \nonumber
&+ \frac{|POS_{\{Age\}}(Walk)_{\mathbb{D}^H_2}|} {|\{x_1, x_2, x_3, x_4\}| + |\{x_5, x_6, x_7\}|} \\ \nonumber
&= \frac{0}{4 + 3}  + \frac{3}{4 + 3}\\ \nonumber
&= \frac{3}{7}
\end{align}

The reason is that, object $x_6$ is in $POS_{\{Age\}}(Walk)_{\mathbb{D}^H_2}$; whereas, if the whole feature $\{Age\}_{\mathbb{D}^H_1}$ and $\{Age\}_{\mathbb{D}^H_2}$ is considered, it is indiscernible with $x_1$ and $x_2$, which would prevent $x_6$ from being a member of $POS_{\{Age\}}(Walk)$ and the final dependency degree would be correct. To overcome this issue, a secure comparison is needed to compare each features' values of each partition with the other one. 

All objects in positive region of each party, should be compared with the objects in the other party to decide on indiscernibilities. In case of any occurrence, numerator of dependency degree should be decreased by one. 

This process starts from ${\mathbb{D}^H_1}$ by checking all objects in $POS_{\{Age\}}(Walk)_{\mathbb{D}^H_1}$; since there is no object in the set, the process proceeds to ${\mathbb{D}^H_2}$. In the second data partition, three objects have been recognized as members of $POS_{\{Age\}}(Walk)_{\mathbb{D}^H_2}$; therefore, this non-empty set leads to the commence of the secure comparison process. The secure comparison of object $x_6$ in ${\mathbb{D}^H_2}$ with the objects $x_1$ and $x_2$ in ${\mathbb{D}^H_1}$ recognizes three objects as indiscernible; therefore, the dependency degree in Equation \ref{2P-HR-Res} should be decreased by one as shown in Equation \ref{2P-HR-Fin1}.

\begin{align}\label{2P-HR-Fin1}
\gamma_{\{Age\}}(Walk) &= \gamma_{\{Age\}}(Walk)_{\mathbb{D}^H_1} \\\nonumber &+ \gamma_{\{Age\}}(Walk)_{\mathbb{D}^H_2} \\ \nonumber&- IND_{\{Age\}}(Walk)_{\mathbb{D}} \\\nonumber
&= \frac{|POS_{\{Age\}}(Walk)_{\mathbb{D}^H_1}|} {|\{x_1, x_2, x_3, x_4\}| + |\{x_5, x_6, x_7\}|} \\\nonumber
&+ \frac{|POS_{\{Age\}}(Walk)_{\mathbb{D}^H_2}|} {|\{x_1, x_2, x_3, x_4\}| + |\{x_5, x_6, x_7\}|} \\\nonumber
&- \frac{1} {|\{x_1, x_2, x_3, x_4\}| + |\{x_5, x_6, x_7\}|} \\\nonumber
&= \frac{0}{4 + 3} + \frac{3}{4 + 3} - \frac{1}{4 + 3} \\\nonumber
&= \frac{2}{7}\\\nonumber
\end{align}
where $IND_{\{Age\}}(Walk)_{\mathbb{D}}$ is the number of indiscernible objects in both partitions.

In order to decide which feature should be selected, the feature selection process continues with calculating the dependency degree of $\{LEMS\}$. The result of applying QuickReduct algorithm on each partition, individually is shown in Equation \ref{2P-HR-QR2}.

\begin{align}\label{2P-HR-QR2}
\nonumber
\gamma_{\{LEMS\}}(Walk)_{\mathbb{D}^H_1} &= \frac{|POS_{\{LEMS\}}(Walk)_{\mathbb{D}^H_1}|} {|\{x_1, x_2, x_3, x_4\}|} \\ \nonumber
&= \frac{|\bigcup_{X \in \{x_1, x_2, x_3, x_4\}/Walk} \underline{LEMS}X|} {|\{x_1, x_2, x_3, x_4\}|}\\ 
&= \frac{|{\{x_1, x_2\}}|} {|\{x_1, x_2, x_3, x_4\}|} = \frac {2}{4}
\end{align}
\begin{align*}
 \nonumber
\gamma_{\{LEMS\}}(Walk)_{\mathbb{D}^H_2} &= \frac{|POS_{\{LEMS\}}(Walk)_{\mathbb{D}^H_2}|} {|\{x_5, x_6, x_7\}|} \\ \nonumber
&= \frac{|\bigcup_{X \in \{x_5, x_6, x_7\}/Walk} \underline{LEMS}X|} {|\{x_5, x_6, x_7\}|}\\ \nonumber
&= \frac{|{\{\}}|} {|\{x_5, x_6, x_7\}|} = \frac {0}{3}
\end{align*}

After calculating the dependency degree of the two parties, number of indiscernible objects should be calculated and subtracted from the final dependency degree. The final result is shown in Equation \ref{2P-HR-Fin2}.

\begin{align}\label{2P-HR-Fin2}
\nonumber
\gamma_{\{LEMS\}}(Walk) &= \gamma_{\{LEMS\}}(Walk)_{\mathbb{D}^H_1} \\\nonumber&+ \gamma_{\{LEMS\}}(Walk)_{\mathbb{D}^H_2} \\\nonumber&- IND_{\{LEMS\}}(Walk)_{\mathbb{D}}\\
&= \frac{|POS_{\{LEMS\}}(Walk)_{\mathbb{D}^H_1}|} {|\{x_1, x_2, x_3, x_4\}| + |\{x_5, x_6, x_7\}|} \\ \nonumber
&+ \frac{|POS_{\{LEMS\}}(Walk)_{\mathbb{D}^H_2}|} {|\{x_1, x_2, x_3, x_4\}| + |\{x_5, x_6, x_7\}|} \\ \nonumber
&- \frac{0} {|\{x_1, x_2, x_3, x_4\}| + |\{x_5, x_6, x_7\}|} \\ \nonumber
&= \frac{2}{4 + 3} +\frac{0}{4 + 3} - \frac{0}{4 + 3}\\ \nonumber
&= \frac{2}{7}
\end{align}

By comparing and selecting feature with the highest dependency degree, the process of feature selection proceeds to the next level by calculating the dependency degree of the new set, which contains $R = \{Age, LEMS\}$ for each parties as shown in Equation \ref{2P-HR-QR3}.
\begin{align}\label{2P-HR-QR3}
\nonumber
\gamma_{\{Age, LEMS\}}(Walk)_{\mathbb{D}^H_1} = \frac{|POS_{\{Age, LEMS\}}(Walk)_{\mathbb{D}^H_1}|} {|\{x_1, x_2, x_3, x_4\}|} \\ \nonumber
= \frac{|\bigcup_{X \in \{x_1, x_2, x_3, x_4\}/Walk} \underline{Age, LEMS}X|} {|\{x_1, x_2, x_3, x_4\}|}\\ \nonumber
= \frac{|{\{x_1, x_2\}}|} {|\{x_1, x_2, x_3, x_4\}|} = \frac {2}{4}\\ 
\\ \nonumber
\gamma_{\{Age, LEMS\}}(Walk)_{\mathbb{D}^H_2} = \frac{|POS_{\{Age, LEMS\}}(Walk)_{\mathbb{D}^H_2}|} {|\{x_5, x_6, x_7\}|} \\ \nonumber
= \frac{|\bigcup_{X \in \{x_5, x_6, x_7\}/Walk} \underline{Age, LEMS}X|} {|\{x_5, x_6, x_7\}|}\\ \nonumber
= \frac{|{\{x_5, x_6, x_7\}}|} {|\{x_5, x_6, x_7\}|} = \frac {3}{3}\\ \nonumber
\end{align}

The final dependency degree for $R = \{Age, LEMS\}$ is calculated and illustrated in Equation \ref{2P-HR-Fin3}.

\begin{align}\label{2P-HR-Fin3}
\nonumber
\gamma_{\{Age, LEMS\}}(Walk) &= \gamma_{\{Age, LEMS\}}(Walk)_{\mathbb{D}^H_1} \\\nonumber&+ \gamma_{\{Age, LEMS\}}(Walk)_{\mathbb{D}^H_2} \\\nonumber&- IND_{\{Age, LEMS\}}(Walk)_{\mathbb{D}}\\
&= \frac{|POS_{\{Age, LEMS\}}(Walk)_{\mathbb{D}^H_1}|} {|\{x_1, x_2, x_3, x_4\}| + |\{x_5, x_6, x_7\}|} \\ \nonumber
&+ \frac{|POS_{\{Age, LEMS\}}(Walk)_{\mathbb{D}^H_2}|} {|\{x_1, x_2, x_3, x_4\}| + |\{x_5, x_6, x_7\}|} \\ \nonumber
&- \frac{0} {|\{x_1, x_2, x_3, x_4\}| + |\{x_5, x_6, x_7\}|} \\ \nonumber
&= \frac{2}{4 + 3} + \frac{3}{4 + 3} - \frac{0}{4 + 3} \\ \nonumber
&= \frac{5}{7}
\end{align}

Based on the greedy nature of QuickReduct algorithm, $R = \{Age, LEMS\}$ is selected, since it ends to the highest dependency degree.

\subsection{Multi parties with horizontally partitioned data (MP-HP)}
In multi-party environments, the most important challenge is to run secure comparison as efficient as possible. Since many parties are involved, each should calculate the dependency degree of each feature in their partitions and also find indiscernible objects. Having a record of indiscernible objects help the whole process by deciding on indiscernibility of objects from other partitions faster. When the secure comparison process is triggered, objects from the other partition are compared with the objects in the indiscernible set of the same partition, initially. If the decision on the indiscernibiliy is finalized, the corresponding dependency degree should be affected. Otherwise, a thorough comparison should be run on all non-indiscernible objects, also.

\subsection{Two parties with vertically partitioned data (2P-VP)}
In case of having vertically partitioned data, three principles should be followed as follows:

\begin{enumerate}
  \item Each partition should have the classification results
  \item Features should have the same order in the whole dataset
  \item A set of indiscernible objects should be created for each partition
\end{enumerate}

As an illustrative example, the dataset in Table \ref{dt} has been partitioned vertically into two datasets and shown in Tables \ref{P1-VP} and \ref{P2-VP}.

\begin{table}[!h]
\caption{The first partition of vertically partitioned data $\mathbb{D}^V_1$}
\centering
\begin{tabular}{ccc}
\hline
\bf{Object} & \bf{Age} & \bf{Walk}\\\hline
$x_1$ & 16-30 & Yes\\
$x_2$ & 16-30 & No\\
$x_3$ & 31-45 & No\\
$x_4$ & 31-45 & Yes\\
$x_5$ & 46-60 & No\\
$x_6$ & 16-30 & Yes\\
$x_7$ & 46-60 & No\\\hline
\end{tabular}
\label{P1-VP}
\end{table}

\begin{table}[!h]
\caption{The second partition of vertically partitioned data $\mathbb{D}^V_2$}
\centering
\begin{tabular}{ccc}
\hline
\bf{Object} & \bf{LEMS} & \bf{Walk}\\\hline
$x_1$ & 50 & Yes\\
$x_2$ & 0 & No\\
$x_3$ & 1-25 & No\\
$x_4$ & 1-25 & Yes\\
$x_5$ & 26-49 & No\\
$x_6$ & 26-49 & Yes\\
$x_7$ & 26-49 & No\\\hline
\end{tabular}
\label{P2-VP}
\end{table}

By referring to the required principles for 2P-VP datasets, each partition has classification outcome and also the order of samples are preserved. The only remaining criterion is sets of indiscernible objects for both parties, which have been calculated and mentioned in Equation \ref{2P-VR-IND}.

\begin{align}\label{2P-VR-IND}
IND_{\{Age\}}(Walk)_{\mathbb{D}^V_1} = \{\{x_1, x_2, x_6\}, \{x_3, x_4\}\} \\
IND_{\{LEMS\}}(Walk)_{\mathbb{D}^V_2} = \{\{x_3, x_4\}, \{x_5, x_6, x_7\}\} \nonumber
\end{align}

As calculated in Equation \ref{QR1}, the dependency degree of each feature in each partition is equal. So, one of feature should be selected to break the tie, since both of them have the same dependency degree. Regardless of which feature is selected, calculating the dependency degree of $R = \{Age, LEMS\}$ requires some efforts. Since all objects are available to each party and the exact value for dependency degree can be calculated, each party should decide on the number of indiscernible objects. 

Party one (or two), needs to know if there is any intersection between the two indiscernible sets, if any, the cardinality of the subset should be subtracted from the number of objects in the dataset. The process is shown in Equation \ref{2P-VR-INDFin}.

\begin{align}\label{2P-VR-INDFin}
\nonumber
IND_{\{Age\}}(Walk)_{\mathbb{D}^V_1} = \{\{x_1, x_2, x_6\}, \{x_3, x_4\}\} \\
IND_{\{LEMS\}}(Walk)_{\mathbb{D}^V_2} = \{\{x_3, x_4\}, \{x_5, x_6, x_7\}\} \\ \nonumber
IND_{\{Age\}}(Walk)_{\mathbb{D}^V_1} \cap IND_{\{LEMS\}}(Walk)_{\mathbb{D}^V_2} \\\nonumber= IND_{\{Age, LEMS\}}(Walk)_{\mathbb{D}} \\ \nonumber
= \{\{x_3, x_4\}\} \nonumber
\end{align}

Therefore, the final dependency degree is illustrated in Equation \ref{2P-VP}. 

\begin{align}\label{2P-VP}
\nonumber
\gamma_{\{Age, LEMS\}}(Walk) &= \frac{7}{7} - \frac{2}{7}\\
&= \frac{5}{7}
\end{align}

In cases which two partitions have more than one feature, each one should calculate the dependency degree of all the features, as well as, their indicernibility sets. Then, a feature with the highest dependency degree should be added to the reduct set. Therefore, there are two cases that should be addressed properly for both partitions, as follows:

\begin{enumerate}
  \item If the selected feature is in the same partition
  \item If the selected feature is in the other partition
\end{enumerate}

For the partition that contains the selected feature, the only task is to build the reduct sets with two members and calculate the dependency degrees without the need of communicating with other party. However, the other partition should have indiscernibility set of the selected features to be able to find the dependency degree of the sets with two members. Hence, a secure comparison should be applied to fulfil this requirement.

\subsection{Multi parties with vertically partitioned data (MP-VP)}
In the environment with more than two parties and vertically partitioned data, the same policy for 2P-VP works fine. The only issue is the amount of communication that is made to/from parties to compute the dependency degrees. Therefore, a computationally inexpensive secure comparison is desired to minimize the overall load.

\section{Conclusion}
Feature selection is the process of selecting important features while discarding the others. This process is usually referred to as a pre-process since it purifies data for a main process, such as classification. Almost all of the previously introduced feature selection methods are not useful for the current needs which involve distributed and decentralized datasets and parties. Therefore, researchers have tried to develop new feature selection methods which can be applied to distributed datasets. In this paper, we have reviewed three feature selection methods, and provided some suggestions to improve their performance when identified. We have also introduced a privacy-preserving feature selection method based on rough set feature selection. The proposed method, has been designed to process both horizontally and vertically partitioned datasets for either two-party or multi-party scenarios. As a future work, we are currently working on privacy-preserving protocols for other standard feature selection methods, as well as their formal security and complexity analyses, along with experimental results on real data. Also, the performance and effectiveness of the proposed method will be examined against UCI datasets. Finally, we will integrate all the proposed protocols in an online privacy-preserving feature selection tool which will be publicly available for non-commercial purposes.

\bibliographystyle{elsarticle-num}
\bibliography{mybib}

\end{document}